# An HST Survey of the Disk Cluster Population of M31. I. WFPC2 Pointings[1]

O. K. Krienke, Seattle Pacific University, Seattle, WA 98119
and
P. W. Hodge, University of Washington, Seattle, WA 98195-1580

**ABSTRACT.** As a follow up to the automated cluster search carried out by Williams and Hodge (2001), we have examined 39 HST WFPC2 pointings to locate and study a comprehensive collection of disk clusters. The Williams technique was effective in finding young clusters, but not intermediate age or old clusters. Our searches have shown that M31 has large numbers of these intermediate and older open clusters, most of them undetected by both the Williams survey and other ground-based searches. We present a catalog of 343 clusters detected on the WFPC images. Extrapolation from our data indicates that the entire disk of M31 contains approximately 80,000 star clusters. We have carried out integrated multi-color photometry of these clusters to ascertain their properties and to compare their properties with cluster systems of other galaxies. We show the cluster luminosity function, the color-magnitude diagram, and the size distribution. Cluster densities and colors show trends with disk position. An age distribution is derived and, though the ages are very uncertain for the fainter clusters, there is evidence for cluster dynamical destruction at about the same rate as for the local Galaxy.

(key words: galaxies:star clusters, galaxies:individual(M31, NGC 224), Local Group)

## 1. INTRODUCTION

### 1.1 Background

The first mention of an "open star cluster" in M31 appears in Hubble's pioneering paper that provided his evidence that M31 is an extragalactic object (Hubble 1929). He suggested that the group of stars cataloged as NGC 206, lying within the southwest section of M31, had properties indicative of its being similar to open clusters in our galaxy. It is now recognized that, while it contains luminous young stars like some open clusters, its size (over 1200 pc) is more similar to very large stellar associations (van den Bergh 1964).

Hodge (1979) used the KPNO 4-m telescope to search for true open clusters. His was a global search, covering all of the M31 disk as it was recognized at that time. The result was a catalog of 403 candidate open clusters, which were primarily young objects, as implied by the fact that they usually appeared resolved on the plates. Subsequently we found that HST images of some of them show that the sample was contaminated by small OB association and asterisms (Williams and Hodge 2001b).

Three-color CCD photometry of a selection of the cataloged clusters was carried out by Hodge, Mateo and Geisler (1987), showing that the clusters sampled are very young objects. More recently Williams and Hodge (2001a) used HST WFPC2 images to obtain CMDs for four young populous ("blue globular") clusters, showing that these objects, while previously classified as globular clusters, have ages of 60-150 million years, thus unlike any traditional globular clusters of our galaxy, as many papers, e. g., Vetesnik (1962), van den Bergh (1967) and many others, had previously suggested on the basis of their colors.

Most recent studies of possible disk clusters in M31 have concentrated on globular-like clusters, determining their abundances, radial velocities, and implied ages. For example, Barmby et al. ( 2000) identified several young objects among globular cluster catalogs. Morrison et al. (2004) measured radial velocities of globular clusters in or projected onto the M31 disk, finding that they could separate disk clusters from halo clusters on the basis of their kinematics. Burstein et al. (2004) used MMT spectra to provide a list of young massive clusters, and Beasley et al. (2004, 2005) noted that several previously-cataloged globular clusters have early-type spectra. Further important work on these objects has been described in papers by Puzia et al. (2005), Fusi-Pecci et al. (2005) and several others. A warning regarding some of this work was published by Cohen et al. (2006), whose Keck survey of four reported "blue globulars" showed that two of these objects are asterisms, recognized only with adaptive optics on ground-based telescopes or by HST.

From experience gained in identifying open clusters in a spiral arm of M31 from HST WFPC2 images (Magnier et al. 1997), we realized that there is a rich population of faint clusters in the disk. Five years ago Williams developed an automated technique to search for small young star clusters in M31, finding 79 such objects on a selection of 13 WFPC2 fields (Williams and Hodge, 2001b). We are now publishing a follow-up to that paper, in which we describe 343clusters identified on 39 WFPC2 pointings, including many clusters too red to have been detected by Williams' technique. This paper reports on our work with HST WFPC2 images; a later paper will report on a study of ACS images.

**1.2 Globular vs open clusters; the terminology problem**

One of the difficulties encountered in the literature of extragalactic star clusters is the confusion that can arise because of the terminology. In the case of the local Galaxy it traditionally has been possible to define globular vs open clusters in terms of mass, kinematics, ages and spatial distribution. However, as first was found for the luminous young clusters in the Magellanic Clouds, for other galaxies this separation is less clear. Currently, some papers refer to young populous clusters as "young globulars" or "blue globulars" and some call them "open clusters", though their structure and appearance are not like most Galactic open clusters. Furthermore the terms "old" and "young" mean different things to different people. In some cases a "young globular" is one that is 8 - 10 Gyr old, a little younger than the traditional 13 Gyr, while another paper might use the term to refer to objects two orders of magnitude younger than that.

For the purposes of this paper, we are adopting the model, suggested by the various recent studies of M31 clusters cited above, that assumes that star clusters in M31 can be separated into two categories according to their kinematics and spatial distribution: halo clusters and disk clusters. We are interested here in disk clusters, regardless of their masses or ages and the results below address their properties on the assumption that all clusters seen within the outlines of the main optical disk are disk clusters, unless radial velocities have established otherwise.

We have carried out this research in an attempt to take advantage of the Hubble Space Telescope's ability to detect a so-far unstudied component of the disk clusters in M31: fainter, smaller, older. We have examined the possibility of determining approximate cluster ages from integrated colors and (for some) CMDs. We also have measured the current cluster formation function and rate and its history and examined the cluster properties as a function of location in M31. A final goal was to determine the cluster destruction rate. Although the data for the oldest and faintest clusters are quite uncertain, it has been possible to produce a first attempt at determining the destruction rate for M31's disk clusters.

## 2. OBSERVATIONAL MATERIAL

For this paper, we searched the archives for all HST WFPC2 pointings that included at least two colors and that were positioned towards the M31 disk. Table 1 lists positions, filters and exposures for the 39 selected pointings.

Figure 1 shows that our sample includes a variety of environments, including bulge positions, as well as locations in arms and interarm regions. The total area covered is 199 arcmin$^2$, which is only approximately 0.44 % of M31's main disk (depending on how its outline is defined).

**2.1 Search techniques**

This paper is a follow-up of the Williams and Hodge (2001b) paper cited above, which used an automatic computer-driven algorithm to identify young clusters by detecting clumped brightness and blue color enhancements in the fields. This technique was found to be effective only for detecting young clusters. Our attempts to find an effective automated way to detect redder and fainter clusters were unsuccessful. Therefore, we used eye searches of the files, done completely independently by each of us. Our discerning criteria were developed and evolved by the large number of experiments that we carried out using artificial clusters, discussed below.

We examined each candidate cluster at two different magnifications, at various levels of brightness and contrast, with images set as both positive and negative and on all available colors, usually B and V or V and I, but in a few cases in three colors. We did not include frames for which there was only one color available in the archives. Both the WC and PC frames were searched.

Figure 2 shows a sample field with the identified clusters indicated.

**2.2 The cluster catalog**

Table 2 lists the 343 detected objects that we consider to be definite clusters. The positions quoted were determined from the HST frames and the headers, using the standard routine, wfpc2_metric, in IDL. As a check on the positions, we compare our data for the few clusters with published positions, finding agreement within a few arcsec. Also we used the same program to measure positions of 30 bright stars on some of the frames and found that our positions agreed with those published by Massey et al. (2006) with a mean difference of 0.2 arcsec.

Table 2 also provides our measurements of the integrated magnitudes and colors of the clusters, together with their uncertainties. In some cases the cluster images were either only partly on the frame or were in the vignetted area of the frame, making the photometry uncertain. These values are identified by having no uncertainties indicated for them. When a cluster was identified but the image was defective, because of bad pixels, column defects or cosmic rays in single exposures, we list position only.

**2.3 Detection efficiencies and completeness**

An extensive array of artificial cluster tests were carried out in order to determine the efficiency of cluster detection as a function of various circumstances, including characteristics of the clusters and of the background against which it must be detected. For this purpose we used six Galactic clusters (the Peaides, h & $\chi$ Persei, NGC 225, IC 2488, NGC 3532, and IC 4651) with known properties, including ages, numbers of stars, magnitudes and colors of stars and sizes, adjusting their characteristics to place them at M31's distance. In addition, we also varied their populations to allow us to test more completely for clusters of various masses for each age. An example of an artificial cluster has been inserted in Figure 2, labeled "AC".

The results of these tests were very instructive. We found that the detectability of clusters depends on several variables. Most clearly is the dependence on luminosity and size. Clusters with absolute magnitudes fainter than V = 22 were difficult to detect even in relatively open fields. Very compact clusters (less than about 3 pc in diameter) were difficult to detect even when bright. And all clusters' detectability was affected in various ways by the nature of the background stellar fields.

Figure 3 shows the results of tests in which clusters of various ages and luminosities were randomly placed in HST M31 fields. The efficiency decreases as the integrated magnitude of the clusters increases; the relationship between cluster magnitude and detectability shows a curve that ranges from near 90% at V = 21 to near 0 at V = 23.5.

We also found that the detectability of clusters depended on the background star density, but in a complicated way. The effect of an enhanced background density reduced the cluster detectability in cases, as in spiral arms, where the background is highly structured. However, for smooth backgrounds, as in the bulge region, the effect was found to enhance the detectability, especially of old clusters with stars of nearly uniform faint brightness. Fig. 3 shows results for three series of tests for various background densities.

A comparison of our list of clusters with that of Williams and Hodge (2001b) shows that 32 of the clusters in Table 2 are common to the earlier catalog. These are indicated in Table 2 with the notation "WH". More than half of the objects found by the automated technique failed our criteria, in all cases because of a lack of central concentration of stars. A review of the images indicates that these objects are all in dense sections of M31's arms, where the luminous blue stars are non-randomly distributed. We consider the discrepant objects to be parts of stellar associations (either physically or statistically clumped), which may or may not eventually survive as stable clusters.

Comparison with the "open clusters" listed by Barmby and Huchra (2001) shows that six of our objects coincide in position with theirs. These are identified in Table 2 by the initials "BH". All of these objects are very clearly young clusters and they are quite massive, judging by their high luminosities. Our search included much fainter clusters. Our use of images of real open clusters, including Galactic clusters with parameters adjusted to M31's distance, as explained above, allowed us to identify low mass clusters with confidence. In this connection, we mention that our original list included an additional 241 objects that we considered candidate clusters, but which were rejected on the basis of comparisons with artificial clusters; some may be clusters, but the evidence is too uncertain. One of these is indicated in Figure 2, labeled "CC".

There are 25 clusters in our list that are known globular clusters. Most of these were the original targets of the HST program. We include them for completeness but exclude them from further discussion, except for those with colors indicating that they are young clusters. Those identified by Sargent et al. (1977) are given their numbers in that catalog, preceded by the letter G. Comparison with the New Bologna Catalog of M31 Globular Clusters (Galleti et al., 2004) shows that 4 objects are in common with that catalog, excluding those in Sargent et al. (1977) and Barmby and Huchra (2001). These clusters are indicated in Table 2 with the initials NBC.

If we assume that our survey is representative of the disk cluster population of M31, then we can calculate the extrapolated total population of clusters of the type identified here, using the percentage of the disk sampled given above. The result is an estimate of a total of 80,000 clusters, a value that is highy dependent on the assumptions made.

## 3. CLUSTER PROPERTIES

We begin this discussion with a few words about reddening. Much of our data was available in only two colors, so direct determinations of cluster reddening was not possible. Though F336W data were available for most frames with B,V, the multi-valued nature of the transformation to Johnson/Cousins U and the filter red-leak (Holtzman et al, 1995) prevented our developing reliable color-color plots to calculate reddening individually. We therefore quote only the measured colors and magnitudes in most of the discussions below. To provide at least a statistical account of the reddening (both from the foreground and internal to M31), we have adopted the average values determined by Williams and Hodge (2001b), which were measured for blue main sequence stars in several disk clusters.

### 3.1 Integrated magnitudes and colors

Integrated magnitudes of the clusters were determined by means of a program written specifically for photometry of small, irregular clusters. Rather than assuming circular symmetry, which is not a good assumption for low-mass, young clusters, our program allowed us to determine magnitudes and colors within a specific outline, which both accounted for the common irregular shape and excluded any bright foreground stars that might contaminate the photometric results. From the HST archives we extracted 12 pointings that used filters F439W and F555W, 8 pointings with F439W, F555W and 814W, and 18 pointings with F555W and F814W.

Photometry was performed with contours taken by cursor at a surface brightness that averaged V = 22.9±0.5/arcsec$^2$. One of the largest sources of error for the photometry is the structured background of the disk field of M31. The uncertainties that we calculate for each cluster's photometry in many cases are dominated by this structure. The errors quoted in this paper include values obtained from the spread in background results from our multiple measures across each field. As in our study of NGC6822 (Krienke & Hodge 2004), 24 adjacent "background" samples of size and shape identical to the cluster were taken. The Chauvenet criterion was applied to these samples, rejecting outliers with less than 0.02 probability of belonging to the desired population, i.e. M31 open cluster background values. Rejections actually took place only at the bright end, involving what are most likely foreground stars, other clusters, or stars of M31 much brighter than the cluster background. Repeated iterations of Chauvenet rejection had little effect on the mean, and a single application was chosen as sufficient. Magnitudes and colors were transformed to the Johnson/Cousins system of B,V,I, (Holtzman 1995).

We note that because we measured the clusters within fairly bright contour levels in order to avoid the effects of a bright and variable background, the listed magnitudes would require aperture corrections to be true integrated magnitudes. These corrections are very small for the majority of our clusters, which are quite faint. Experiments indicate that the corrections would be on the order of 0.03 magnitudes at V = 22.5, increasing to 0.4 magnitudes for bright globular clusters at V = 18.

**3.2 Photometric Errors**

Figure 4 shows the distribution of photometric errors for the clusters for each of the colors. The errors are small, averaging 0.04 mag in V, for example, for clusters with V < 20 and they increase to an average value as large as 0.20 mag for the faintest clusters. We note that more precise values for these clusters will not be easy to obtain, as the unknown value for a cluster's background produces a statistical uncertainty that could only be overcome by very deep photometry of individual stars and star-by-star background subtraction.

For a check on our photometruc system, we measured 13 stars located on two of the M31 frames and compared them to the photometry of Massey, et al.( 2006), with agreement to within ± 0.04 in V, ± 0.07 in B, and ± 0.05 in I.

**3.3 Comparisons with previous photometry**

Almost all of our clusters are so faint that they have no previous published photometry available. However, for a few of the brighter clusters, especially the globular clusters that in some cases were the reason for the pointings, we have compared our photometry with that of others. Because we chose to measure clusters with fairly small apertures in order to minimize the effects of background contamination on the colors, our values for the integrated magnitudes of the clusters are systematically fainter than most of the others. The colors, on the other hand, agree well within the quoted uncertainties.

## 4. CLUSTER PROPERTIES

**4.1 The luminosity function**

Figure 5 gives our results for the luminosity function for the cluster sample. The bright end of the function includes the classical, old globular clusters, most of which are probably not disk clusters. The brightest non-globular clusters (the "blue globulars") have absolute magnitudes of M(V) ~ -8, while the faintest clusters in the sample are at V ~ -1. Note that the number of the faintest clusters is highly uncertain because of the effects of the detection efficiency. We point out that our data, though uncertain at these levels, reaches much fainter, smaller mass clusters, than are typically sampled in external galaxies. As described in Section 8, an intermediate-age cluster with an absolute magnitude of M(V) = -1 has a mass of only approximately 100 solar masses.

The shape of the luminosity function is essentially identical to that determined for other star-forming galaxies, such as the Magellanic Clouds (Bica et al 1996) and the sample studied by Boutloukos and Lamers (2003). As discussed in Section 8, the luminosity function shape, as well as the color-magnitude diagram, is determined by the formation function, the formation history, evolutionary fading and dynamical destruction.

**4.2 The color magnitude diagram**

Figures 6a and 6b show the observed color-magnitude diagrams of the clusters in the survey. These diagrams, while not readily interpretable in terms of detailed physical processes because of the multiple parameters that enter into their structure, provide a useful index of trends and can be compared to similar diagrams for other galaxies to detect similarities or differences. Note that we have plotted the observed colors and magnitudes.

As is shown most clearly in Figure 6a, most of the clusters are fairly blue, making up a vertical "main sequence" with colors between B –V = 0.0 and B – V = 0.6. Less thickly populated is a red region which extends to B – V = 1.5 and in which the magnitudes are weighted to the faint end, except for a few very bright clusters, especially conspicuous in Figure 6b. The latter are the classical globular clusters, which are overrepresented in our sample because of being specific targets for the HST pointings.

The few clusters that lie beyond these limits include, on the blue side, some very young objects, probably less than 10 million years old (e.g., see Girardi et al. 2005). The clusters redder than B – V = 1.5 (and V – I = 2) are either highly-reddened objects or objects whose colors are distorted by large background variations.

There are very few published samples of star clusters in other galaxies that extend to faint luminosities. The most complete collections of photometry of such clusters are available for the Magellanic Clouds (e.g., see Bica et al., 1991, and the reviews by van den Bergh, 2000, and by Westerlund, 1997) and for NGC 6822 (Krienke and Hodge 2003). The general distributions of objects in the color-magnitude diagrams of clusters in these galaxies are fairly similar. The bimodal distribution of clusters in the LMC CMD is not conspicuous in the M31 cluster diagram except within the top three magnitudes, where the red "clump" is artificially enhanced by the presence of classical globulars, as explained above. Note also that our survey reaches clusters that are three magnitudes fainter than those measured for the LMC; thus except for the tiny population of clusters in NGC 6822, our data are sampling a new regime for extragalactic cluster populations.

**4.3 The size distribution**

We did not plan to determine accurate sizes for the clusters, but we do have some useful information about their sizes that was a byproduct of the photometry. Our program recorded the width and height of the chosen area for the photometry and, although this does not give a precisely defined size, it does give a good indication of the approximate dimensions of the cluster. Figure 7 shows the distribution of the diameter values measured in the x direction on the frames. The maximum of the distribution is at 11 pc, while the largest clusters are nearly 36 pc across. The smallest measured objects are 3 pc in size (just under 1 arcsec). Figure 7 is not corrected for detection efficiency and there are probably many very small clusters that were missed.

Although our data are not more than estimates and thus should not be used for any quantitative comparisons, the referee has requested that we make a rough comparison with other cluster samples. For this we have to compare to papers that use diameters similarly-made as eye estimates, rather than what would be preferred, e.g., sizes based on profile-fitting. Therefore, in Figure 7 we show for comparison data for open clusters in the LMC, based on ground-based eye-estimates (Hodge 1988). The distributions are similar, except for a displacement along the x axis, which is probably partly the result of the fact that the M31 data are at best upper limits. It also suggests that many small clusters may have been missed by our searches because they are not well resolved at M31, as we also conclude from our experiments with faint artificial clusters (Section 2.3).

## 5. SPATIAL DISTRIBUTIONS

**5.1 Global radial distribution**

Figure 8 displays the global distribution of cluster densities as a function of radius in the disk; the numbers of clusters per pointing are plotted (the deprojected area of a pointing is 1.041 square kiloparsecs). It was assumed that all clusters are in the flat disk and that the angle of inclination of the disk is 12.5 degrees. The diagram shows that the cluster density reaches its greatest values in the well-known active arm areas at radial distances of 6-12 kpc, but that the dispersion in densities is large throughout the range of 6-13 kpc. Inside 6 kpc the cluster density appears to be constant at approximately 5 clusters/pointing (0.00154 clusters per square pc). Beyond 13 kpc the cluster density steeply decreases to zero.

Knowing the colors of the clusters allows us to deduce something about the nature of the density variations in terms of the formation history of clusters in different regions. Assuming that the disk clusters have nearly circular orbits, so that the radial distance from the center remains unchanged, we interpret the mean colors at the different radial distances as a population indicator, where redder means indicate a preponderance of old clusters and bluer clusters the reverse. Figure 9 shows the mean colors for clusters in each pointing as a function of their radial distances for those for which data is available in B and V. For this as well as for the V – I data, the display looks like a scatter diagram, suggesting that there is a large dispersion in age throughout the disk. The existence of an excess of very blue clusters at distances of near 6 kpc indicates that there has been a recent concentration of cluster formation at those distances, as is well-known for star formation from other evidence (van den Bergh 2000).

**5.2 Arm vs interarm clusters**

A second examination of cluster density trends can be made by measuring the deprojected distance of clusters from the nearest spiral arm or arm segment. We have used the GALAX image of M31 (Wyder 2005) to establish the positions of spiral arms. We created a high-contrast version of the published image, transformed to face-on using an inclination angle of 12.5 degrees. We plot in Figure 10 the deprojected cluster densities as a function of the shortest distance to a spiral arm or arm fragment. There is a strong tendency for clusters to lie close to arms, in agreement with the distribution of stars. In interpreting the diagram one should note the fact that the average distance between spiral arms in M31 is about 5 kpc, so that the only clusters farther than that from an arm are those in the outer areas, where the arms are unusually widely-spaced.

Figure 11 shows the mean (B – V) colors for clusters in the various pointings as a function of distance from the nearest arm or arm fragment. The diagram shows a very rough trend in the sense that clusters nearer the arms tend to be younger on the average than those farther. This is consistent with the idea that young clusters are preferentially formed in arms and then drift away as the arm density wave passes through.

## 6. THE CLUSTER AGE DISTRIBUTION

The cluster sample is a potentially valuable source of information on the age distribution of clusters in the M31 disk. As the first attempt to find and measure a complete sample of such clusters, it can be used to make some preliminary conclusions about the age distribution. These conclusions must be only preliminary, however, because the sample suffers from several uncertainties, most of them the result of the bright and varied background of the galaxy disk, which affects both the photometry and the detection efficiency. The following is a list of the principle difficulties.

1. Photometric uncertainties are a problem, though they can at least be evaluated, as discussed in Section 3 above. As we note there, repeated measurements do not necessarily lead to better values, as the uneven background presents an intrinsic uncertainty that can only be overcome by using a much larger space telescope in order to have photometry of the individual stars in the clusters.

2. Abundance uncertainties also degrade conclusions about the ages of the clusters. When sufficient spectroscopy of M31 stars of known ages has established the age-metallicity relation (if there is one in general), this uncertainty can be reduced.

3. The reddening of the individual clusters is unknown. We have attempted to determine reddenings for those clusters for which there are frames in three colors (e.g., U, B and V), but found that the uncertainties, especially in the transformation to U magnitudes, are too large to find reliable reddenings.

4. The models themselves are uncertain, in the sense that different theorists find significantly different relations between integrated colors and ages for single-age populations (see, for example, the papers in Chavez and Valls-Gabaud, 2006). We have adopted the models of Girardi et al. (2005), but would have found somewhat different ages had we used other choices.

5. For very low mass clusters there are stochastic effects caused by the discrete number of the brightest stars.

Keeping in mind these limitations, we have calculated approximate ages for all of our clusters. Using the adopted mean reddening and binning the data, we have derived an age-frequency diagram for our clusters (Figure 12). This diagram shows a steeply decreasing frequency of clusters with increasing age. A least squares linear fit to the data gives

$$\text{Log}(N) = 1.198 \log t + 8.769,$$

where N is the number of clusters per million years and t is the age in years. These diagrams are based on the measured sample only, uncorrected for detection efficiency.

A similar decrease is found for other cluster populations, such as those in the Milky Way Galaxy (Kharchenko et a. 2005, Lamers et al. 2005), in NGC 6822 (Krienke and Hodge 2004) and in several other galaxies (Boutloukos and Lamers 2003, Gieles et al. 2006).

There are at least three factors that can lead to a decrease in cluster numbers with age such as that shown in Figure 12. First is the fading of cluster brightness due to stellar evolution. This leads to a dropping out of low mass clusters with age as their brightness drops below the search magnitude limit. We have calculated the fading effect based on the Girardi models for solar metallicity. Clusters of 100 solar mass will fade to integrated magnitudes below our limit when they are about 600 million years old, while clusters 13 billion years old will be undetectable if their masses are smaller than about 2500 solar masses (we have used these data, assuming that $Z = 0.018$ for the clusters in the absence of a more precise idea regarding their chemical compositions). Of course, the numbers of clusters lost at each magnitude is modulated by the detection efficiencies given in Section 2.3.

Second, clusters can lose mass as they evolve because of strong stellar winds and supernovae. The rate of mass loss is especially significant for small mass clusters. An analytical expression for this mass loss rate is given by Lamers et al. (2005b).

Third, clusters are subject to tidal destruction and have a finite lifetime in a galaxy's disk. Cluster destruction was first discussed by Oort (1958) and Spitzer (1958) and has been the subject of numerous recent studies, based on analytical, n-body and observational analyses (see papers by Lamers and Gieles (2006), Gieles et al. (2006), Baumgardt and Makino (2003) and the many references therein). Mass loss from clusters can occur because of the general galactic tidal field, because of tidal shocking, because of giant molecular cloud encounters, and because of internal dynamical evolution ("cluster evaporation"). It is generally believed that the giant molecular cloud encounters are the principle cause of cluster destruction in the solar neighborhood (Lamers and Gieles 2006), but the situation may be different in other environments.

Making some simple assumptions, it is possible to disentangle the various effects that determine the shape of Figure 12. The most important assumption is that the rate of formation and the mass function for clusters has been uniform since M31 formed. (This assumption can eventually be checked when it is possible to determine the detailed star formation history of the M31 disk, making the reasonable but unproven assumption that star and cluster formation occurred in parallel). The true age-frequency relation for clusters can then be determined by using the color-age relation and by taking into account the detection efficiencies and the evolutionary fading.

The age distribution for the cluster sample can be derived from the colors, the assumed reddenings, the Girardi color-age relations, the detection efficiencies and the evolutionary fading. The combined uncertainties, ignoring the uncertainties in the assumption of a uniform formation history, are very large. A least squares linear fit to the data in Figure 13 leads to the relation

$$\log(N) = -0.691 \log(t) + 5.526,$$

where the quantities are defined above.

However, we point out that the data are not yet good enough even to determine whether a linear fit is best. The last data point has an especially large uncertainty because of its heavy dependence on several uncertain quantities. If omitted from the diagram, the best fit is curved and shows a similarity to that derived for the clusters near the sun (Lamers and Gieles 2006). They are compared in Figure 13. At least tentatively, we conclude that the disk clusters in M31 have relatively short lifetimes, similar to those derived more precisely for the Milky Way Galaxy's disk clusters.

## 7. SUMMARY

A survey of disk clusters in M31 on HST WFPC2 images has led to a catalog of 343 clusters. These are generlly both fainter and younger than previously-detected clusters in M31 and providethe first source of information on the cluster population for low-mass star clusters in the galaxy. The luminosity function for clusters is derived to absolute magnitudes of $M(V) = -1$. It is an increasing function with decreasing luminosity, and our data do not detect a turnover at the faintest magnitudes. The spatial densities of clusters in the disk are nearly constant from the central bulge out to a radial distance of 6 kpc and then it shows a large dispersion in the active area that ranges from 7 to 14 kpc. The relationship of cluster density and age with distance from spiral arm structures follows expectations: the clusters peak in number near the arms and those nearest the arms tend to be younger on average. In an attempt to measure the age-frequency relation for clusters, we find tentative evidence that the cluster destruction rate in the M31 disk is similar to that in our local Galaxy.

We are grateful to the American Astronomical Society's Small Grant Program for publication funds and to Henny Lamers and Anil Seth for many helpful suggestions and ideas.

# REFERENCES

Barmby, P., Huchra, J. P., Brodie, J. P., Forbes, D. A., Schroder, L. L., Grillmair, C. J. 2000, AJ, 119, 727

Barmby, P. and Huchra, J. 2001, AJ, 122, 2458

Baumgardt, H. and Makino, J. 2003, MNRAS, 340, 227

Beasley, M. A., Brodie, J. A., Strader, J., Forbes, D. A., Proctor, R. N., Barmby, P. and Huchra, J. 2004, AJ, 128, 1623

Beasley, M. A., Brodie, J. A., Strader, J., Forbes, D. A., Proctor, R. N., Barmby, P. and Huchra, J. 2005, AJ, 129, 1412

Bica, E., Claria, J. J., Dottori, H., Santos, J. F. C., and Piatti, A. E. 1996, ApJS, 102, 57

Boutloukos, S.G. and Lamers, H. J., G., L., M. 2003, MNRAS, 338,717

Burstein, D., Li, Y., Freeman, K. C., et al. 2004, ApJ, 614, 158

Chavez, M. and Valls-Gabaud, D. (eds.) 2006, Resolved Stellar Populations (San Francisco: ASP Conf. Series)

Cohen, J. G., Matthews, K. and Cameron, P. B. 2005 ApJ, 634L, 45

Fusi-Pecci, F., Bellazzini, M., Buzzoni, A. de Simone, E. Federici, L. and Galleti, S. 2005, AJ, 130, 554

Gieles, M., Portegies Zwart, S. F., Baumgardt, H. et al. 2006, MNRAS, in press.

Girardi, L. 2006, http://pleiadi.pd.astro.it

Hodge, P. W. 1979, AJ, 84, 744

Hodge, P. W. 1988, PASP, 100, 1051

Hodge, P. W.; Mateo, M.; Lee, M. G.; Geisler, D. 1987, PASP, 99, 173

Holtzman, J. A.; Burrows, C. J.; Casertano, S.; Hester, J. J.; Trauger, J. T.; Watson, A. M.; Worthey, G. 1995, PASP, 107, 1065

Hubble, E. P. 1929, ApJ, 69, 99

Kharchenko, N., Piskunov, A., Roeser,S. Schilbach, E. and Scholz, R.-D. 2005, A&A, 438, 1163

Krienke, O. K. & Hodge, P. W. 2004, PASP, 116, 497

Lamers, H. J. G. L. M., Gieles, M., Bastien, N., et al. 2005 A&A, 441, 117

Lamers, H. J. G. L. M. and Gieles, M. 2006, A&A, in press

Ma, J., Zhou, X., Burstein, D, Yang, Y., Fan, Z., Chen, J., Jiang, Z., Wu, Z., Wu,, J. and Zhang, T. 2006, A&A 449, 143

Magnier, E. A., Hodge, P. W., Battinelli, P., Lewin, W. H. G. and van Paradijs, J. 1997, MNRAS, 292, 490

Massey, P., Olsen, K. A. G., Hodge, P. W., Strong, Shay B., Jacoby, G. H., Schlingman, W., Smith, R. C. 2006, AJ, 131, 2478

Morrison, H. L., Harding, P., Perrett, K. and Hurley-Keller, D. 2004, ApJ, 603, 87

Oort, J. 1958, Ricerche Astronomiche, 5, 507

Puzia, T. H., Perrett, K. M. and Bridges, T. J. 2005, A&A, 434, 909

Sargent, W., Kowal, C., Hartwick, F. and van den Bergh, S. 1977, AJ, 82, 947

Spitzer, L. J. 1958, ApJ 127, 17

van den Bergh, S. 1964, ApJS, 9, 65

van den Bergh, S. 1969, ApJS, 19, 145

van den Bergh, S. 2000, The Local Group of Galaxies (Cambridge: Cambridge Univ. Press)

Vetesnik, M. 1962, Bull. Astron. Inst. Czech., 13, 180

Westerlund, B.1997, The Magellanic Clouds (Cambridge: Cambridge Univ. Press)

Williams, B. F. and Hodge, P. W. 2001a, ApJ 548, 190

Williams, B. F. and Hodge, P. W. 2001b, ApJ, 559, 851

Wyder, T. K. 2005, BAAS 207, 5203

TABLE 1

HST WFPC2 POINTINGS SEARCHED

| DataSet | $\alpha_{2000}$ | | | $\delta_{2000}$ | | | Filter | | | |
|---|---|---|---|---|---|---|---|---|---|---|
| | | | | | | | F336W exposure (s) | F439W exposure (s) | F555W exposure (s) | F814W exposure (s) |
| U4CA5501R | 0 | 36 | 19.0 | +40 | 53 | 17.6 | | | 5300 | 5400 |
| U4CA0701R | 0 | 39 | 32.2 | +40 | 30 | 48.1 | | | 5300 | 5400 |
| U5BJ0101R | 0 | 39 | 47.4 | +40 | 31 | 58.0 | 3600 | 1600 | 1200 | |
| U5BJ0201R | 0 | 40 | 1.58 | +40 | 34 | 14.8 | 3600 | 1600 | 1200 | |
| U8MG0109M | 0 | 40 | 3.1 | +40 | 45 | 39.8 | 1050 | 1200 | 300 | 600 |

Notes: Table 1 is published in its entirety in the electronic version of the *PASP*. Portions are shown here for guidance regarding form and content. Datasets are identified by the first entry in the HST archive, regardless of filter.  [a]Dataset U2DG0107T used F450W

TABLE 2
STAR CLUSTERS DETECTED IN M31 WFPC2 HST POINTINGS

| Name | $\alpha_{2000}$ | | | $\delta_{2000}$ | | | V | Verr | B-V | B-Verr | V-I | V-Ierr |
|---|---|---|---|---|---|---|---|---|---|---|---|---|
| SECTION A: PREVIOUSLY IDENTIFIED CLUSTERS ||||||||||||
| G11 | 0 | 36 | 19.95 | +40 | 53 | 29.0 | 16.35 | 0.06 | | | 0.95 | 0.06 |
| G33 | 0 | 39 | 32.98 | +40 | 31 | 3.2 | 15.56 | 0.01 | | | 1.17 | 0.02 |
| G38 | 0 | 39 | 48.06 | +40 | 31 | 42.6 | 16.44 | 0.03 | 0.33 | 0.04 | | |
| WH | 0 | 39 | 52.42 | +40 | 31 | 41.2 | 20.25 | 0.08 | 0.16 | 0.08 | | |
| WH | 0 | 40 | 0.02 | +40 | 33 | 26.1 | 19.29 | 0.07 | 0.24 | 0.08 | | |
| SECTION B: CLUSTERS IDENTIFIED IN THIS STUDY ||||||||||||
| KHM 31-1 | 0 | 36 | 19.33 | +40 | 52 | 53.3 | 24.62 | 0.26 | | | 1.01 | 0.43 |
| KHM 31-2 | 0 | 39 | 30.19 | +40 | 31 | 30.0 | 22.01 | 0.01 | | | 1.27 | 0.24 |
| KHM 31-3 | 0 | 39 | 32.95 | +40 | 31 | 7.4 | 21.95 | 0.01 | | | 1.04 | 0.37 |
| KHM 31-4 | 0 | 39 | 39.45 | +40 | 31 | 30.6 | 21.94 | 0.37 | 0.98 | 0.41 | | |
| KHM 31-5 | 0 | 39 | 40.54 | +40 | 31 | 53.5 | 19.97 | 0.06 | 0.48 | 0.07 | | |

Notes: Table 2 is published in its entirety in the electronic version of the *PASP*. Two portions are shown here for guidance regarding form and content. Clusters with no photometric errors reported are in the vignetted regions of the HST frames, so the photometry is only approximate. For defective images only position is listed.

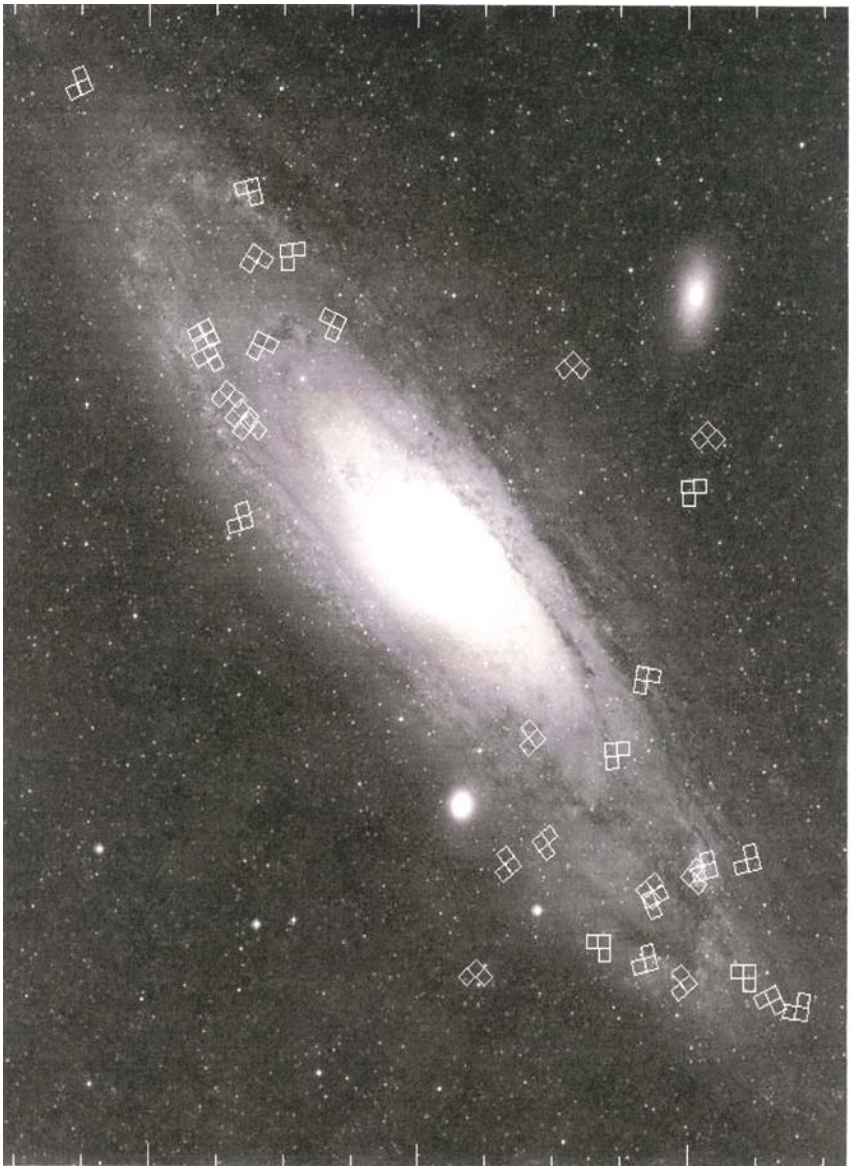

Figure 1 The locations of the HST WFPC2 fields in M31 that we searched for disk star clusters. The photograph was obtained with the Palomar Observatory's 48-inch Schmidt telescope (as it was called then) by one of the authors in 1960. The pointings U4CA5501R and U2830201T lie outside this photograph to the west and to the south, respectively. The field shown is 85 arc min wide.

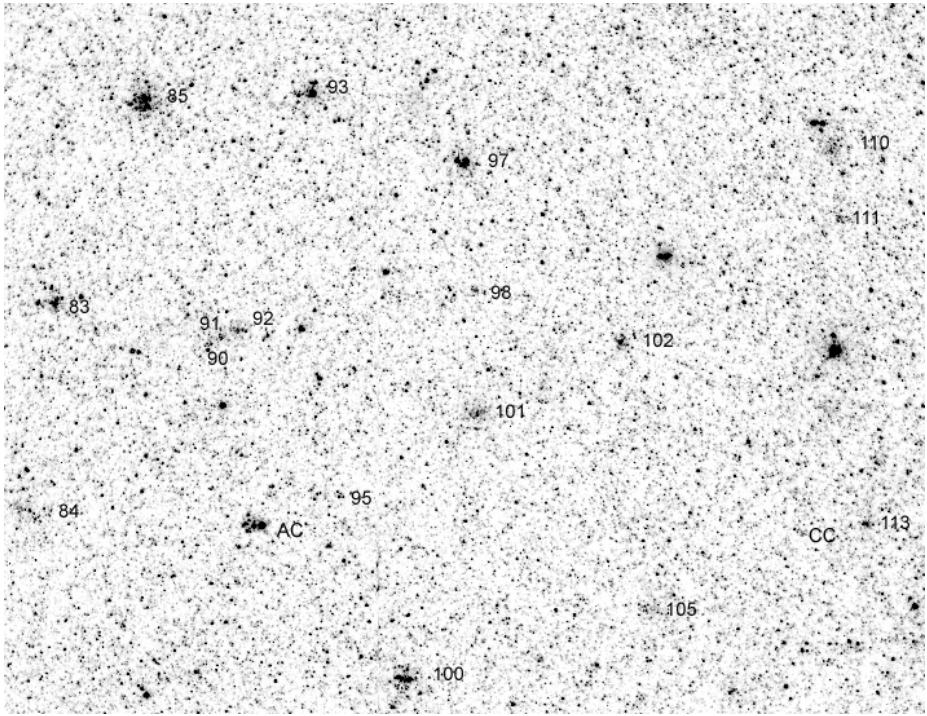

Figure 2 A sample field in a spiral arm region of M31 with the identified clusters marked. A cluster candidate that is an example of the 241 objects considered possible but too uncertain to be included is indicated by "CC". An example of one of the artificial clusters used is indicated by "AC". The field of view is 70.5 arcsec wide.

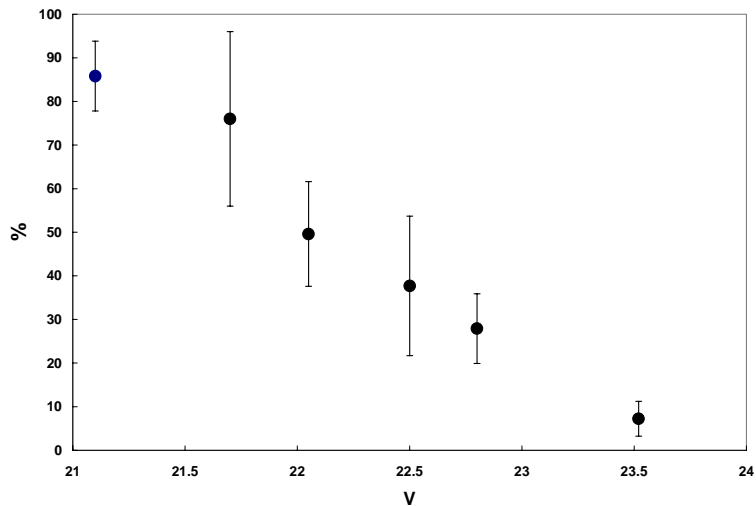

Figure 3 The results of artificial cluster tests for the completeness of the cluster searches in fields with various background densities and with artificial clusters with various characteristics. The error bars in the figure represent the root mean square error for the set of experiments.

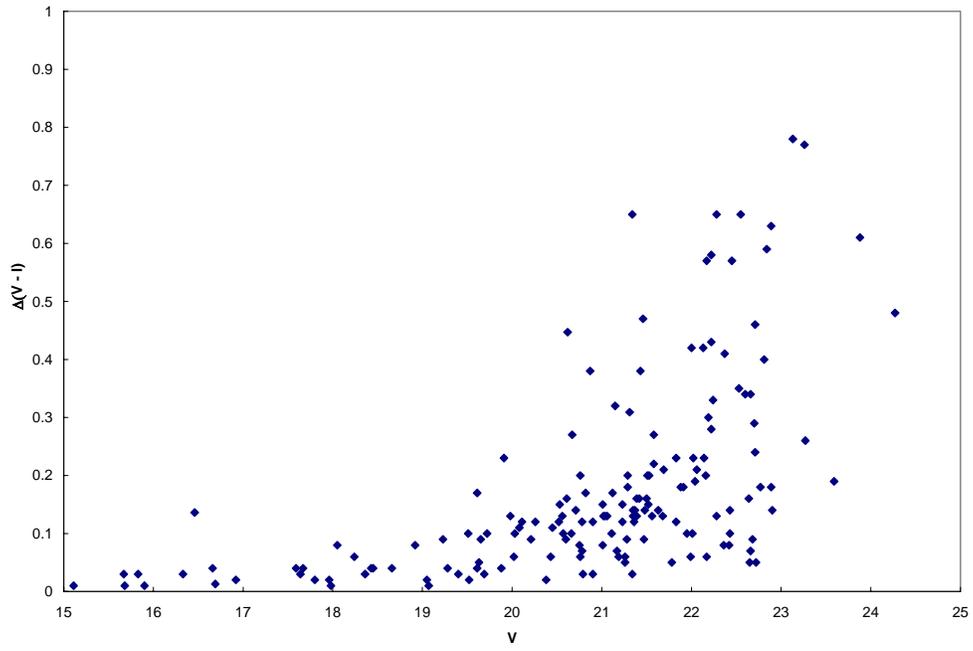

Figure 4a  Measurement errors as a function of magnitude and color.

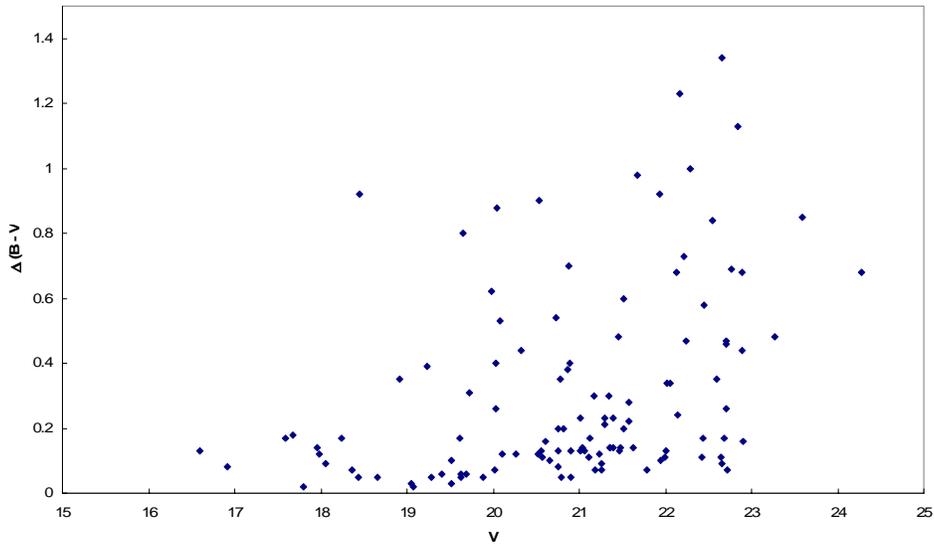

Figure 4b

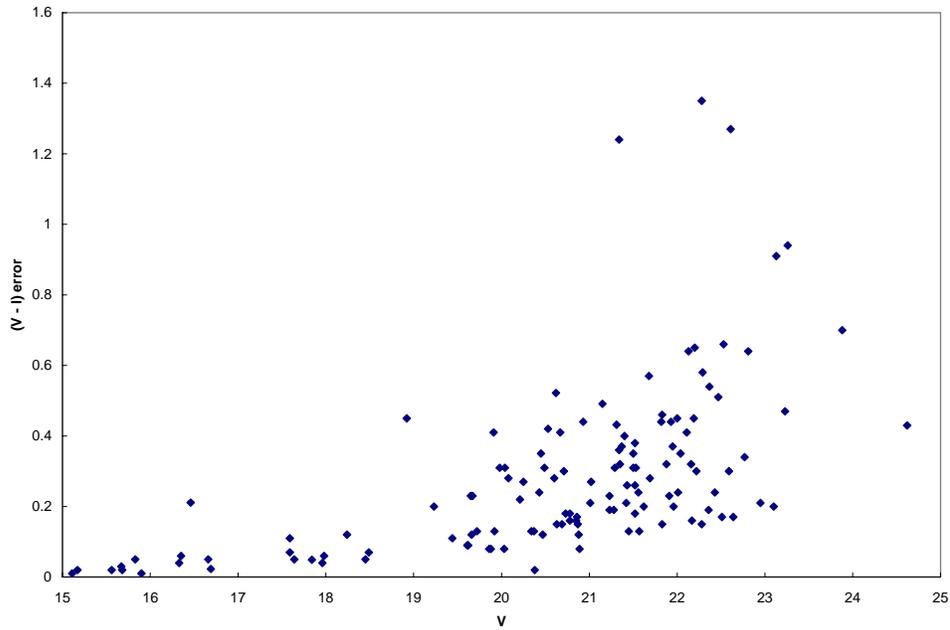

Figure 4c

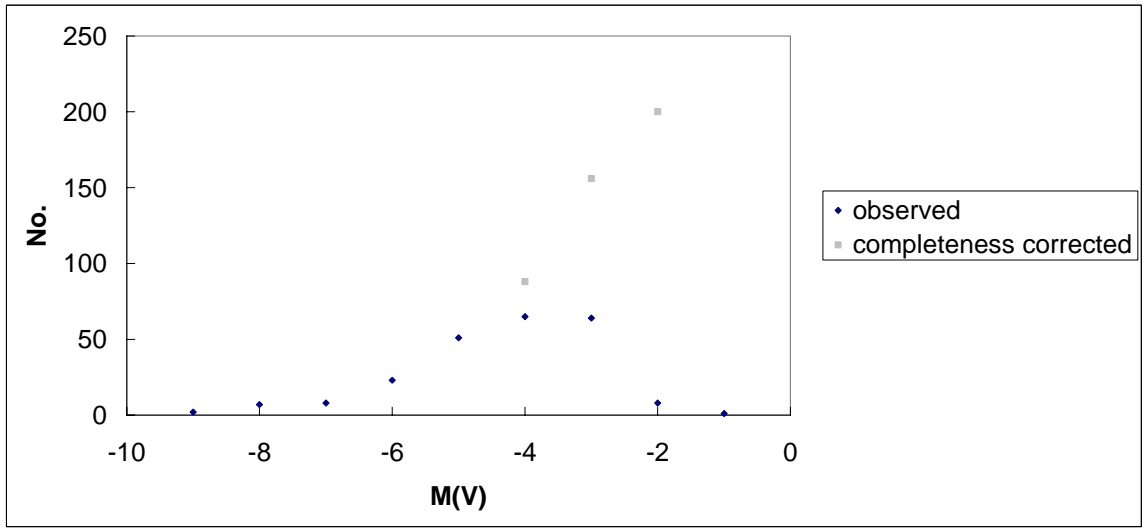

Figure 5  The observed and completeness-corrected cluster luminosity function. We note that our values for the absolute magnitudes are 0.0-0.2 magnitudes fainter than limiting magnitudes (see text).

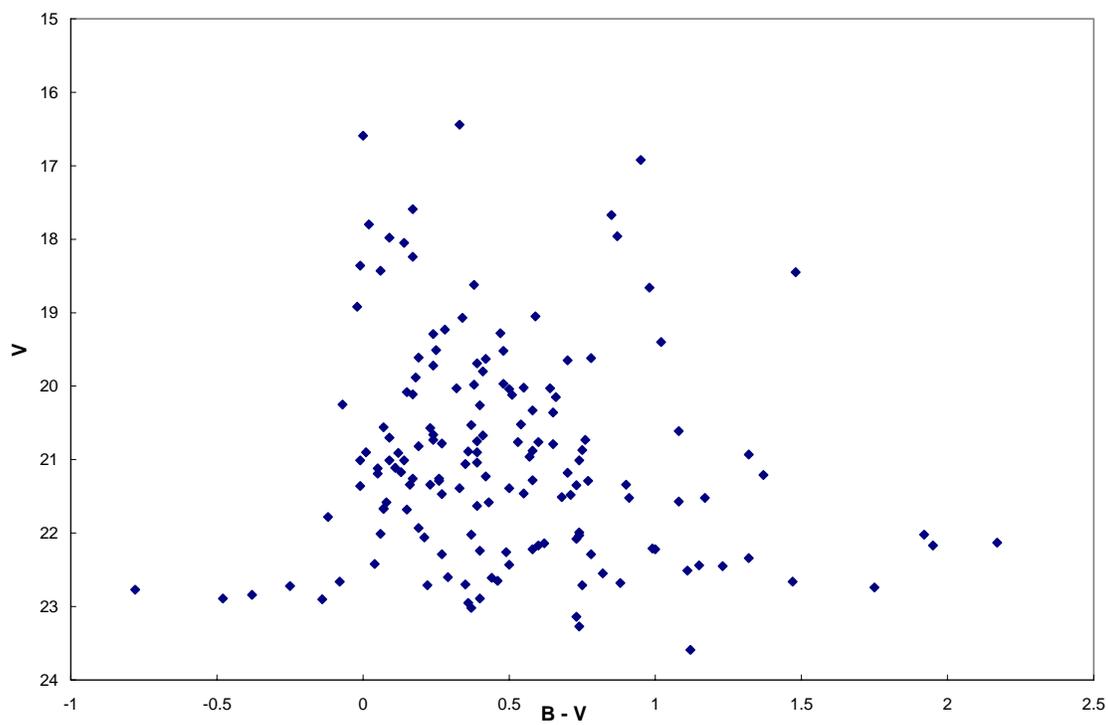

Figure 6a The observed cluster color-magnitude diagrams

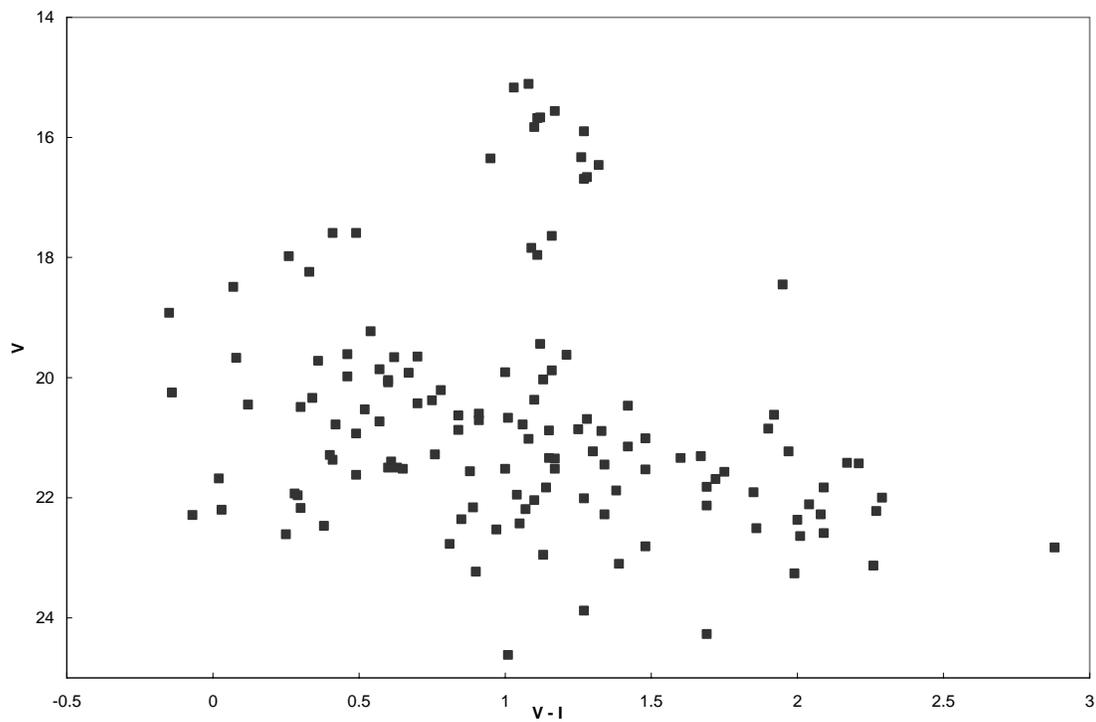

Figure 6b

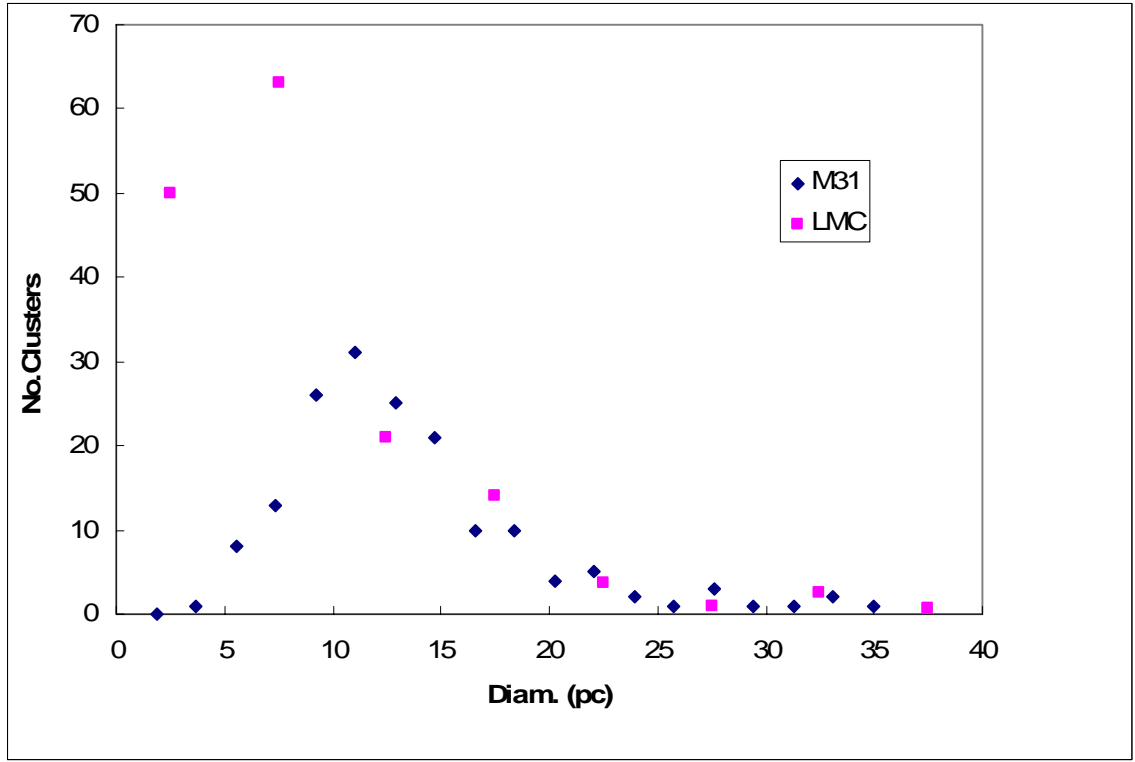

Figure 7 The approximate cluster size distribution copared to that for the LMC (see text for reservations about the scale).

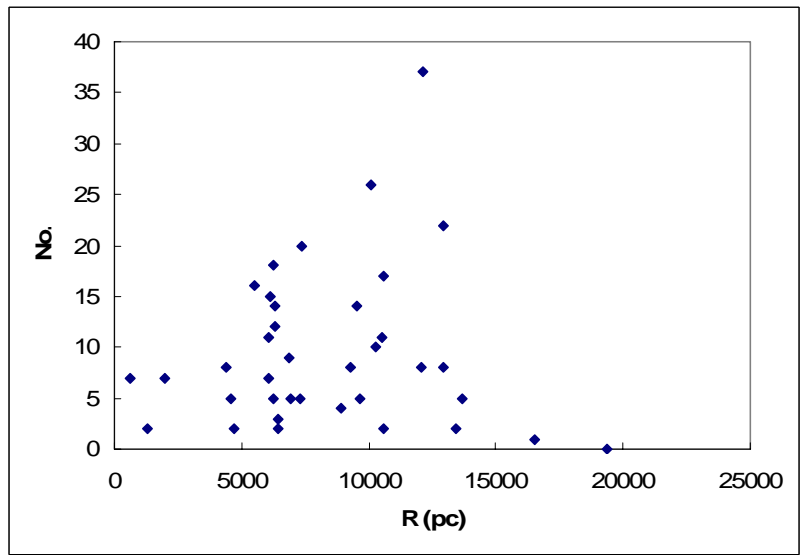

Figure 8 The number of clusters per pointing as a function of R, the distance from the center of M31, assuming that all clusters lie in the plane.

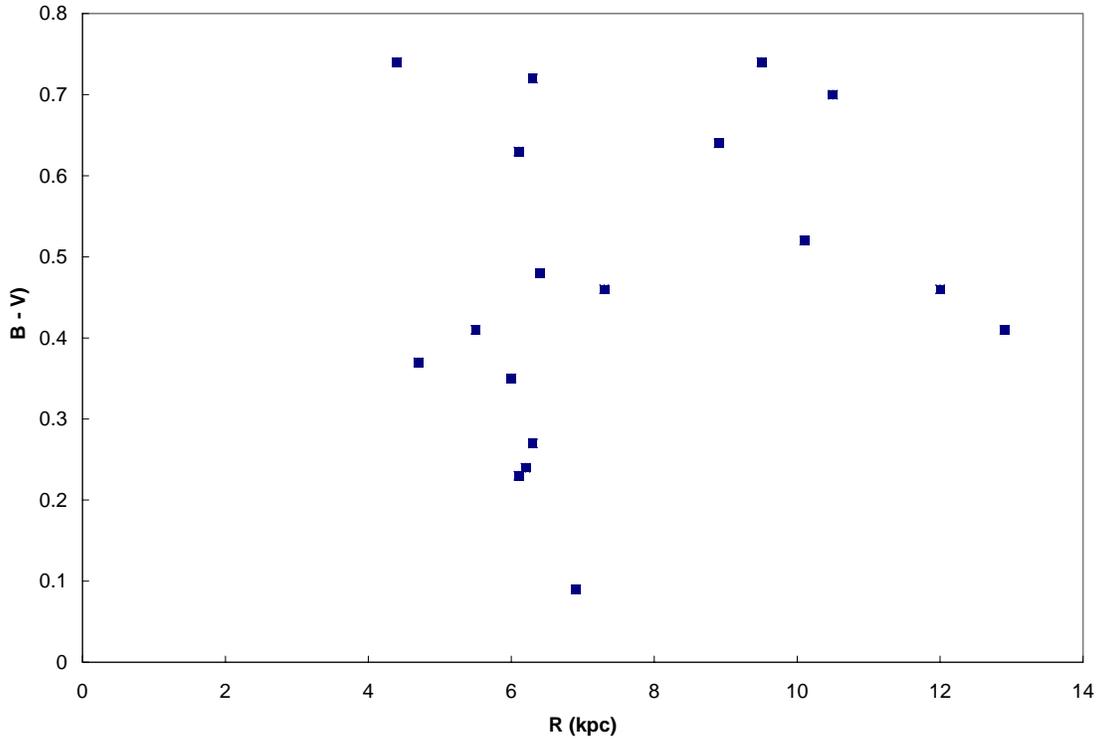

Figure 9  Mean colors of clusters in a pointing as a function of R, the distance from the nucleus, for pointings with B and V data. A similar diagram can be made for V, I data.

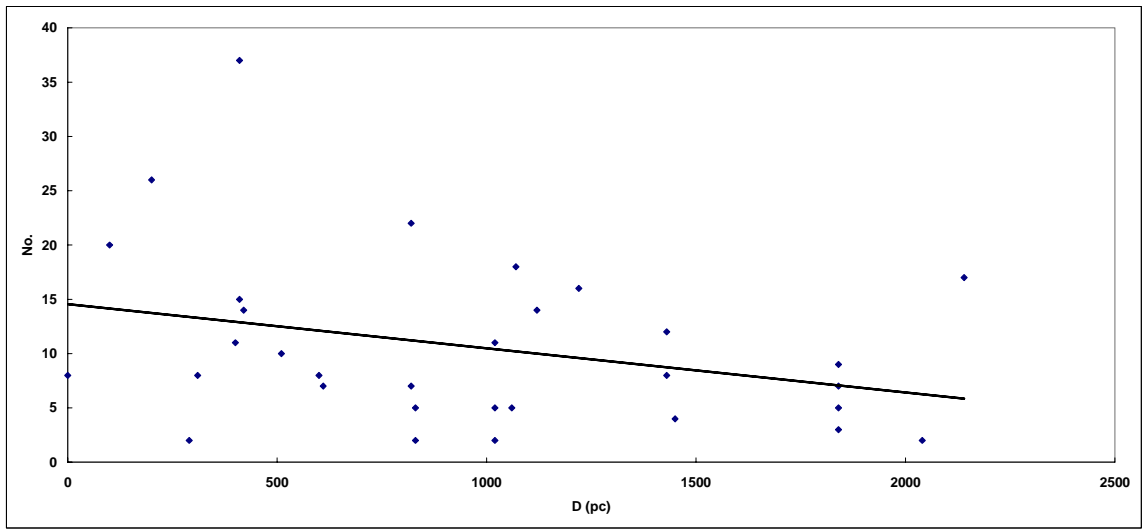

Figure 10  The number of clusters per pointing as a function of D, the distance in the plane from the nearest star-forming area. The line is a least-squares linear fit.

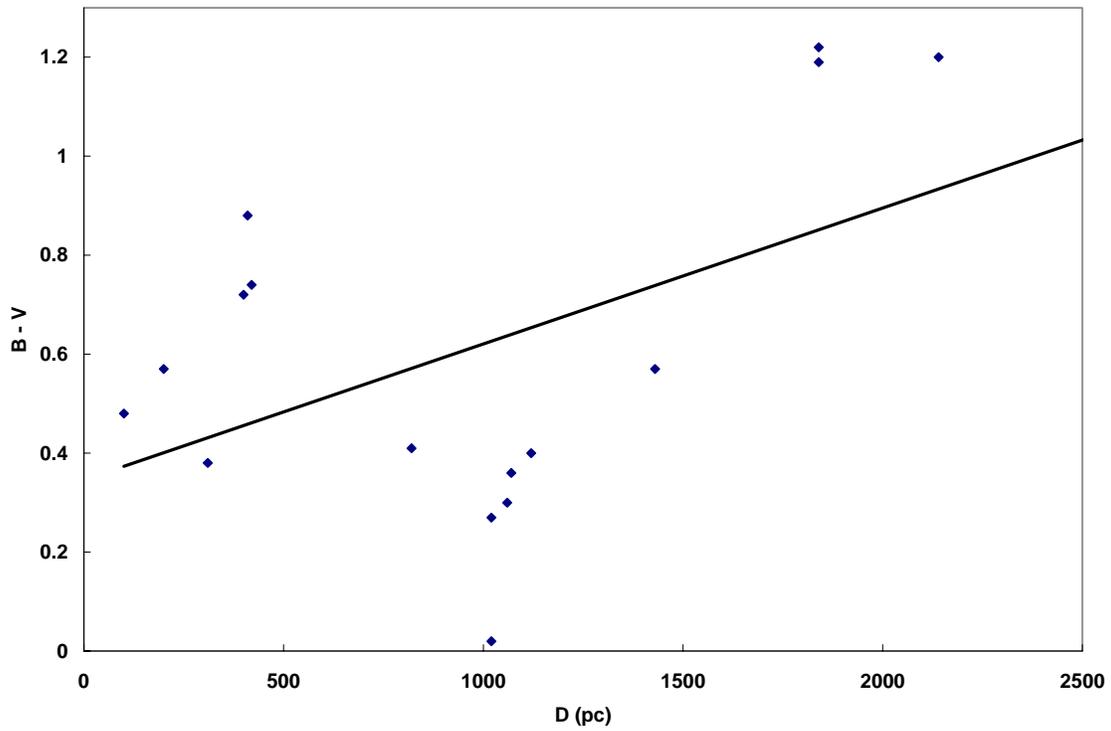

Figure 11 The mean of the cluster colors vs D, distance from the nearest star-forming region. The line is a least-squares linear fit.

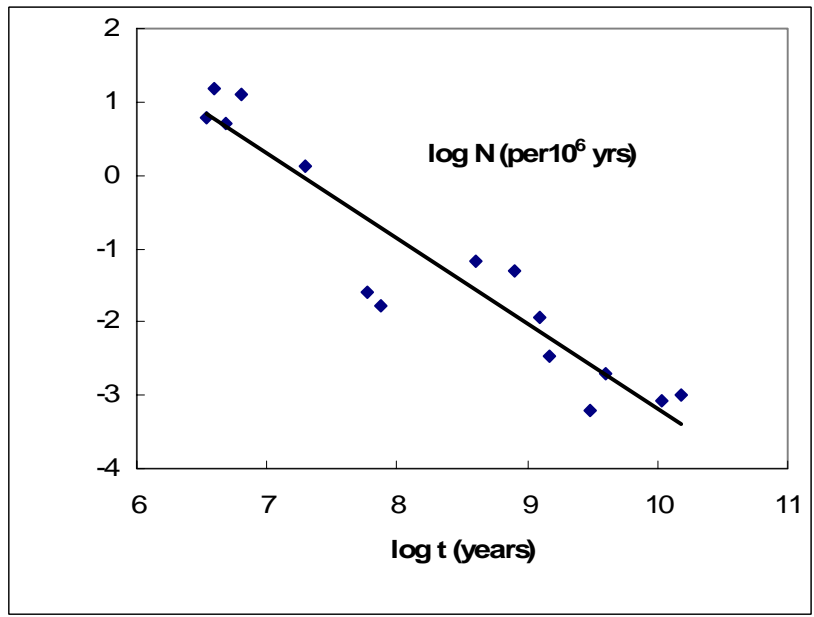

Figure 12 The age distribution for clusters, based on their colors, corrected statistically for reddening. The line is a least-squares linear fit.

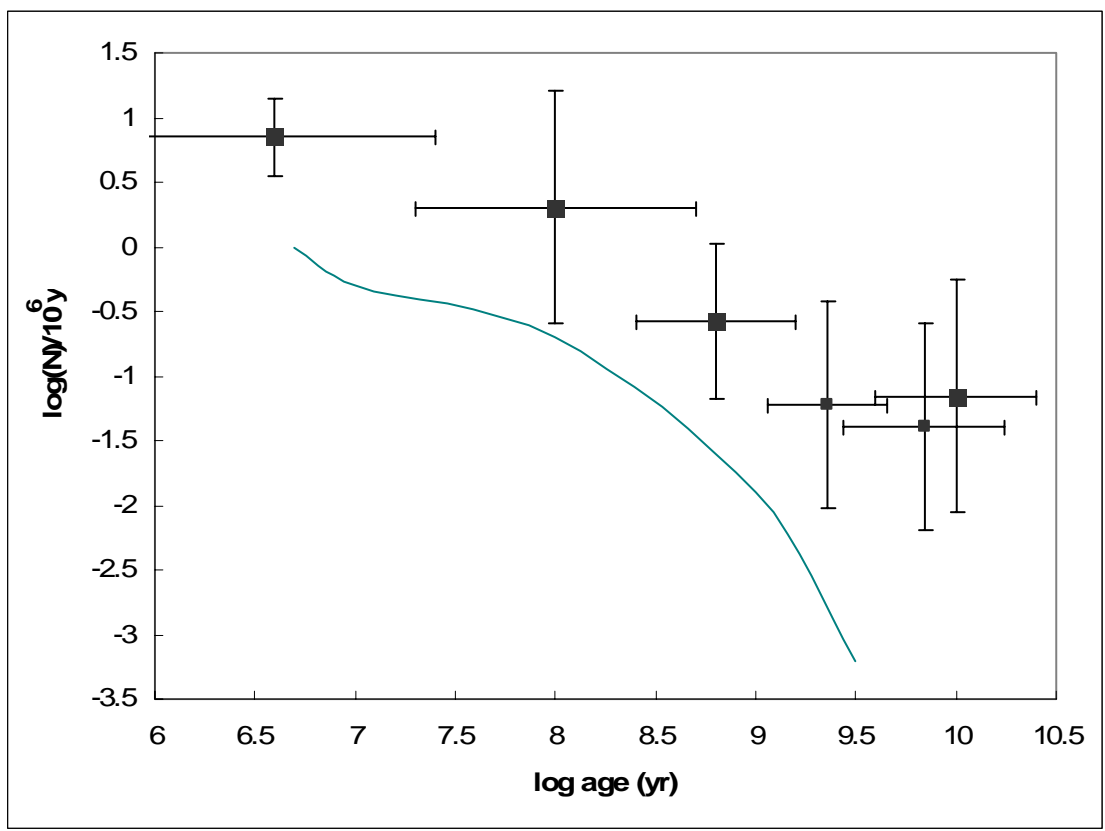

Figure 13 An approximate age-frequency diagram for disk clusters in M31, corrected for evolutionary fading and for detection efficiency (points with error bars). The line is the age-frequency diagram for clusters in the solar neighborhood (Lamers and Gieles 2006), arbitrarily shifted along the y axis for comparison purposes.